\documentclass[12pt]{article}
\usepackage{amsmath}
\usepackage{graphicx}
\begin{document}
\baselineskip=18 pt
\begin{center}
{\large{\bf  A Type D Non-Vacuum Spacetime with Causality Violating Curves, and Its Physical Interpretation}}
\end{center}

\vspace{.5cm}

\begin{center}

{\bf Faizuddin Ahmed}\footnote{faizuddinahmed15@gmail.com},\\ 
{\it Ajmal College of Arts and Science, Dhubri-783324, Assam, India}
\end{center}

\vspace{.5cm}

\begin{abstract}

We present a topologically trivial, non-vacuum solution of the Einstein's field equations in four-dimensions, which is regular everywhere. The metric admits circular closed timelike curves, which appear beyond the null curve, and these timelike curves are linearly stable under linear perturbations. Additionally, the spacetime admits null geodesics curve which are not closed, and the metric is of type D in the Petrov classification scheme. The stress-energy tensor anisotropic fluid satisfy the different energy conditions and a generalization of Equation-of-State parameter of perfect fluid $p=\omega\,\rho$. The metric admits a twisting, shearfree, non-exapnding timelike geodesic congruence. Finally, the physical interpretation of this solution, based on the study of the equation of the geodesics deviation, will be presented.

\end{abstract}

\vspace{0.3cm}


\vspace{.5cm}

\section{Introduction}

The Einstein field equations of Genereal Relativity are the set of non-linear partial differential equations whose exact solution is very hard. Some well-known solutions of the field equations admit Closed Causal Curves (CCCs) in the form of closed timelike curves (CTCs), closed timelike geodesics (CTGs) and closed null geodesics (CNGs). The presence of such curves in a spacetime violates the causality condition. Examples of these spacetime are the G{\"o}del's Cosmological solution \cite{Go}, van Stockum solution \cite{Sto}, Tipler's rotating cylinder \cite{Tip}, traversable wormholes \cite{MTY,Morri}, and the wrap dripe models \cite{Alcu,lobo} violate the weak energy condition (WEC), Gott's solution \cite{Gott}, Krasnikov spacetime \cite{Kras}, electrovac spacetime \cite{Stead}, and pure radiation field spacetimes \cite{Faiz,Faiz2,Faiz3,Faiz4} have CTCs. Some well-known vacuum spacetimes such as the Kerr and Kerr-Newmann black holes solution \cite{Kerr,Cart} (see also \cite{Hawking}), NUT-Taub metric \cite{NUT}, Bonnor metric \cite{Bon1,Bon2}, Ori metric \cite{Ori1}, locally isometric AdS metric \cite{Faiz5}, and type N Einstein spacetime \cite{Faiz6} have CTCs. In addition, some other CTC spacetimes possesses a naked singularity ({\it e.g.} \cite{Sar2,Faiz7,Faiz8,Faiz9}. Hawking proposed a Chronology Protection Conjecture \cite{Haw} which states that the laws of physics will always prevent a spacetime to form CTCs. However, the general proof of Chronology protection conjecture has not yet existed.

\section{The spacetime with divergence-free curvature}

Consider the following line element in $(t,x,y,z)$ coordinates 
\begin{equation}
ds^2=-dt^2+dx^2+\left(1-\alpha_{0}^2\,x^2\right)dy^2-2\,\alpha_0\,x\,dt\,dy+dz^2,
\label{1}
\end{equation}
where $\alpha_0>0$ is a real number. The metric is topologically trivial, and the ranges of the coordinates are 
\begin{equation}
-\infty < t < \infty,\quad -\infty < x < \infty,\quad -\infty < y < \infty,\quad -\infty < z < \infty.
\label{range}
\end{equation}
The metric has signature $(-,+,+,+)$ and the determinant of the corresponding metric tensor $g_{\mu\nu}$ is
\begin{equation}
det\;g=-1,
\label{2}
\end{equation}
which is regular everywhere even at $x=0$. The non-zero components of the Einstein tensor $G^{\mu\nu}$ are
\begin{equation}
G^{t}_{t}=-\frac{3\,\alpha_{0}^2}{4},\quad G^{t}_{y}=-\alpha_{0}^3\,x,\quad G^{x}_{x}=G^{y}_{y}=-G^{z}_{z}=\frac{\alpha_{0}^2}{4}.
\label{3}
\end{equation}
The scalar curvature invariants of the spacetime
\begin{equation}
R=\frac{\alpha_{0}^2}{2},\quad R^{\mu\nu}\,R_{\mu\nu}=\frac{3}{4}\,\alpha_{0}^4,\quad R^{\mu\nu\rho\sigma}\,R_{\mu\nu\rho\sigma}=\frac{11}{4}\,\alpha_{0}^4,
\label{4}
\end{equation}
are non-vanishing constant. Therefore, the presented spacetime is free from curvature divergence.

\subsection{Stress-energy tensor and the kinematic parameters}

We consider the stress-energy tensor anisotropic fluid for the metric (\ref{1}) given by
\begin{equation}
T_{\mu\nu}=(\rho+p_{y})\,U_{\mu}\,U_{\nu}+p_{y}\,g_{\mu\nu}+(p_{x}-p_{y})\,\eta_{\mu}\,\eta_{\nu}+(p_{z}-p_{y})\,\zeta_{\mu}\,\zeta_{\nu},
\label{5}
\end{equation}
where $\rho$ as the energy density, $p_{x}$, $p_{y}$, and $p_{z}$ are pressures. Here $U^{\mu}$ is the timelike unit four-velocity vector, $\eta_{\mu}$ and $\zeta_{\mu}$ are the spacelike unit vector along $x$ and $z$ direction, respectively. For the metric (\ref{1}), these are defined by  
\begin{eqnarray}
U^{\mu}&=&\delta^{\mu}_{t},\quad \eta^{\mu}=\delta^{\mu}_{x},\quad \zeta^{\mu}=\delta^{\mu}_{z},\quad U^{\mu}\,U_{\mu}=-1,\nonumber\\
\eta^{\mu}\,\eta_{\mu}&=&1=\zeta^{\mu}\,\zeta_{\mu},\quad U^{\mu}\,\eta_{\mu}=0=U^{\mu}\,\zeta_{\mu}=\eta^{\mu}\,\zeta_{\mu}.
\label{6}
\end{eqnarray}
The non-zero components of the stress-energy tensor (\ref{5}) using (\ref{6}) are
\begin{equation}
T^{t}_{t}=-\rho,\quad T^{t}_{y}=-\alpha\,x\,(\rho+p_{y}),\quad T^{x}_{x}=p_{x},\quad T^{y}_{y}=p_{y},\quad T^{z}_{z}=p_{z}.
\label{7}
\end{equation}
And its trace given by
\begin{equation}
T^{\mu}_{\,\mu}=T=-\rho+p_{x}+p_{y}+p_{z}.
\label{8}
\end{equation}

The Einstein's field equations (taking cosmological constant $\Lambda=0$) are given by
\begin{equation}
G_{\mu\nu}=T_{\mu\nu},\quad \mu,\nu=0,1,2,3,
\label{9}
\end{equation}
where $G_{\mu\nu}$ is the Einstein tensor. Here units are chosen such that $c=1$ and $8\,\pi\,G=1$. Equating the field equations (\ref{9}) using (\ref{3}) and (\ref{7}) we get
\begin{eqnarray}
\rho&=&\frac{3}{4}\,\alpha_{0}^{2},\quad p_{x}=\frac{1}{4}\,\alpha_{0}^{2},\quad p_{y}=\frac{1}{4}\,\alpha_{0}^{2},\quad p_{z}=-\frac{1}{4}\,\alpha_{0}^{2},\quad T=-R.
\label{10}
\end{eqnarray}
The matter-energy source satisfy the different energy conditions \cite{Hawking}.

The above stress-energy tensor may be a generalization of Equation-of-State (EoS) parameter of perfect fluid by taking EoS parameter separately on each spatial axis. The stress-energy tensor of perfect fluid is given by
\begin{equation}
T^{j}_{i}=\mbox{diag}\,\left[ T^{0}_{0}, T^{1}_{1}, T^{2}_{2}, T^{3}_{3} \right]=\mbox{diag}\,\left[-\rho, p, p, p\right].
\label{diag}
\end{equation}
We parametrize it as follows :
\begin{eqnarray}
T^{j}_{i}=\mbox{diag}\,[ -\rho, p_{x}, p_{y}, p_{z} ]&=&\mbox{diag}\,[-1, \omega_{x}, \omega_{y}, \omega_{z} ]\,\rho\nonumber\\
&=&\mbox{diag}\,[-1, \omega, \omega, (\omega+\delta) ]\,\rho,
\label{diag2}
\end{eqnarray}
where $\omega_{x}$, $\omega_{y}$ and $\omega_{z}$ are the directional EoS parameters along the $x$, $y$, and $z$ axis, respectively. Here $\omega$ is the deviation-free EoS parameter of the perfect fluid. We have parameterized the deviation from isotropy by setting $\omega_{x}=\omega$, $\omega_{y}=\omega$ and then introducing skewness parameter $\delta$ that is the deviation from $\omega$ along the $z$.

From the stress-energy tensor form (\ref{diag2}) using (\ref{10}), one will get the Equation-of-state as radiation type
\begin{equation}
p=\omega\,\rho,\quad \omega=\frac{1}{3},\quad \delta=-\frac{2}{3},
\label{12}
\end{equation}
where $p$ as the isotropic pressure of the fluid. Therefore, our stress-energy tensor is a generalization of Equation-of-State (EoS) parameter of perfect fluid (radiation type), where $\omega=\frac{1}{3}$ is the deviation-free EoS parameter and $\delta=-\frac{2}{3}$ is the deviation parameter from isotropy along the $z$-direction.

The kinematic parameters, the {\it expansion} $\Theta$, the {\it acceleration} vector $\dot{U}^{\mu}$, the {\it shear tensor} $\sigma_{\mu\nu}$ and the {\it vorticity tensor} $\omega_{\mu\nu}$ associated with the fluid four velocity-vector are defined by
\begin{eqnarray}
\boldsymbol{\Theta}&=&U^{\mu}_{\,\,;\,\mu},\nonumber\\
\boldsymbol{a}^{\mu}&=&\dot{U}^{\mu}=U^{\mu\,;\,\nu}\,U_{\nu},\nonumber\\
\boldsymbol{\sigma}_{\mu\nu}&=&U_{(\mu\,;\,\nu)}+\dot{U}_{(\mu}\,U_{\nu)}-\frac{1}{3}\,\boldsymbol{\Theta}\,h_{\mu\nu},\nonumber\\
\boldsymbol{\omega}_{\mu\nu}&=&U_{[\mu\,;\,\nu]}+\dot{U}_{[\mu}\,U_{\nu]},
\label{13}
\end{eqnarray}
where $h_{\mu\nu}=g_{\mu\nu}+U_{\mu}\,U_{\nu}$ is the projection tensor. For the spacetime (\ref{1}), these parameter have the following expression
\begin{equation}
\boldsymbol{\Theta}=0,\quad \boldsymbol{\sigma}_{\mu\nu}=0,\quad \boldsymbol{\omega}_{xy}=-\boldsymbol{\omega}_{xy}=\frac{1}{2}\,\alpha_0 \quad \mbox{and} \quad \boldsymbol{a}^{\mu}=0.
\label{14}
\end{equation}
The magnitude of vorticity tensor $\boldsymbol{\omega}_{\mu\nu}$, $\boldsymbol{\omega}=\sqrt{\frac{1}{2}\,{\boldsymbol{\omega}}^{\mu\nu}\,{\boldsymbol{\omega}}_{\mu\nu}}=\frac{1}{2}\,\alpha_0$.

\section{Closed Timelike Curves of the spacetime}

The presented spacetime admit circular closed timelike curves which appears beyond the null curve.

Consider a closed curve $\gamma$ defined by $t=t_0$, $x=x_0$ and $z=z_0$, where $t_0$, $x_0$, $z_0$ are constants. Here the $y$ coordinate is chosen periodic, that is each $y$ identified $y+y_0$ for a certain parameter $y_0>0$ (see \cite{Ori1}). From the metric (\ref{1}), we get
\begin{equation}
ds^2=\left(1-\alpha_{0}^2\,x^2\right)\,dy^2.
\label{15}
\end{equation}
These curves are null curve provided $ds^2=0$ for $x^2=x_{0}^2$, spacelike provided $ds^2>0$ for $x^2<x_{0}^2$, but become timelike provided $ds^2<0$ for $x^2>x_{0}^2$. Therefore, the closed curves defined by $t=t_0$, $x>x_0$, and $z=z_0$ being timelike, are closed timelike curves. Thus the formation of closed timelike curves take places beyond the null curve, {\it i.e.,} in the region satisfying $x^2>x_{0}^2$, where $x_{0}^2=\frac{1}{\alpha_{0}^2}$. These curves evolve from an initial spacelike hypersurface. For that we calculate the norm of the vector $\nabla_{\mu} t$ (or by determining the sign of the component $g^{tt}$ in the metric tensor $g^{\mu\nu}$) \cite{Faiz4}. From the metric (\ref{1}), we get
\begin{equation}
g^{tt}=-\left(1-\alpha_{0}^2\,x^2\right).
\label{16}
\end{equation}
A hypersurface $t=const$ is spacelike provided $g^{tt}<0$ for $x^2<x_{0}^2$, but become timelike provided $g^{tt}>0$ for $x^2>x_{0}^2$, and null curve $x^2=x_{0}^2$ serve as the Chronology horizon. Thus, spacelike $t=const$ hypersurface can be choosen as initial conditions over which the initial data may be specified. Therefore, the formation of closed timelike curves here is identical to the metric in \cite{Go,Faiz,Faiz4}.

\subsection{Stability of closed timelike curves}

To analyze the stability of CTCs, we used the method adopted in \cite{Faiz,Faiz3,Faiz4,Rosa1}. The closed timelike curves considered here have the parametric form
\begin{equation}
t=t_0,\quad x^2=x^{2}_{*}>x^{2}_0\quad,\quad y\sim y+y_0\quad \mbox{and} \quad z=z_0,
\label{a1}
\end{equation}
where $t_0$, $x_{*}$, $z_0$ are constants.

A CTC $\gamma$ satisfies the system of equations
\begin{equation}
\ddot{x}^{\mu}+\Gamma^{\mu}_{\alpha\beta}\dot{x}^{\alpha}\dot{x}^{\beta}=a^{\mu}(x),
\label{a2}
\end{equation}
where the dot indicates derivative w. r. t. proper time and $a^{\mu}$ is a four acceleration.

We consider a small perturbation $\bar{x}^{\mu}=x^{\mu}+\xi^{\mu}$ in (\ref{a2}). After perturbation of the system of equations, one can obtain a set of differential equation satisfied by the perturbation $\xi$. We find that 
\begin{eqnarray}
\ddot{\xi}^0&+&k^{2}_{0}\,\xi^0+k_{1}\,\dot{\xi}^1+k_{0}\,\dot{\xi}^3=0,\nonumber\\
\ddot{\xi}^1&-&k_2\,\dot{\xi}^0=0,\nonumber\\
\ddot{\xi}^2&=&0,\nonumber\\
\ddot{\xi}^3&+&(k_0+k_3)\,\dot{\xi}^0=0.
\label{b1}
\end{eqnarray}
Here 
\begin{eqnarray}
k_0&=&\alpha_0\,\dot{y},\quad k_1=2\,\alpha_{0}^2\,x_{*}\,\dot{y}=2\,n\,k_0,\quad k_2=\alpha_{0}^2\,x_{*}\,\dot{y}=n\,k_0=\frac{k_1}{2},\nonumber\\
k_3&=&\alpha_{0}^3\,x^{2}_{*}\,\dot{y}=n^2\,k_0, \quad x_{*}=\frac{n}{\alpha_0},\quad n>1.
\label{dd}
\end{eqnarray}

The set of differential equations above can be solved exactly and we obtain the following set of solutions
\begin{eqnarray}
\xi^0(s)&=&A-\frac{c_2}{\omega}\,\cos (\omega\,s)+\frac{c_1}{\omega}\,\sin (\omega\,s),\nonumber\\
\xi^1(s)&=&c_3+s\,c_4-\frac{1}{\omega}\,[c_1\,\cos (\omega\,s)+c_2\,\sin (\omega\,s)],\nonumber\\
\xi^2(s)&=&c_5+s\,c_6,\\
\label{mm11}
\xi^4(s)&=&c_7+s\,c_8+\frac{(1+n^2)}{n\,\omega}\,[c_1\,\cos (\omega\,s)+c_2\,\sin (\omega\,s)]\nonumber,
\end{eqnarray}
where $c_i$, $i=1,\dots,8$ are constants of integration, and $\omega=n\,k_0$. The above solutions satisfy the differential equations (\ref{b1}) provided $A=-\frac{2\,n^2}{\omega}\,c_4$. For simplicity, we choose here $c_4=0$ so that $A=0$. Therefore, the set of solutions are
\begin{eqnarray}
\xi^0(s)&=&\frac{1}{\omega}\,[c_1\,\sin (\omega\,s)-c_2\,\cos (\omega\,s)],\nonumber\\
\xi^1(s)&=&c_3-\frac{1}{\omega}\,[c_1\,\cos (\omega\,s)+c_2\,\sin (\omega\,s)],\nonumber\\
\xi^2(s)&=&c_5+s\,c_6,\nonumber\\
\xi^4(s)&=&c_7+s\,c_8+\frac{(1+n^2)}{n\,\omega}\,[c_1\,\cos (\omega\,s)+c_2\,(\sin \omega\,s)],
\label{mm22}
\end{eqnarray}

To establish the stability of these orbits (\ref{mm22}), one can calculate the largest invariant Lyapunov exponent, a measure of  stability of these curves which we discussed in \cite{Faiz,Faiz3,Faiz4}. In our case here, we find the Lyapunov exponent defined by 
\begin{eqnarray}
\Lambda&=&\lim_{s \rightarrow \infty}\frac{1}{s}\log \frac{\|{\xi} (s)\|}{\|{\xi} (0)\|}=\lim_{s \rightarrow \infty}\frac{1}{2\,s}\log \frac{\|{\xi} (s)\|^2}{\|{\xi} (0)\|^2}\nonumber\\
&=& \lim_{s \rightarrow \infty}\frac{1}{2\,s}\log {\|{\xi} (s)\|}^2-\lim_{s \rightarrow \infty}\frac{1}{2\,s}\log {\|{\xi} (0)\|}^2\nonumber\\
&=&0,
\label{b2}
\end{eqnarray}
which indicates stablility of these curves under small linear perturbations, where the Riemannian norm is defined by $\|{\xi} (s)\|=\sqrt{|g_{\mu\nu}\,\xi^{\mu}\,\xi^{\nu}|}$. Therefore, the closed timelike curves of the spacetime discussed earlier are stable under linear perturbation.

\subsection{Parametric curves of the metric : closed timelike curves}

For the metric (\ref{1}), we choose two set of parametric curves defined by
\begin{eqnarray}
t(s)&=&c_1+c_2\,\sin (b\,s),\nonumber\\
x(s)&=&c_3+c_4\,\sin (b\,s),\nonumber\\
y(s)&=&c_5+c_6\,s,\nonumber\\
z(s)&=&c_7+c_8\,\cos (b\,s),
\label{17}
\end{eqnarray}
\begin{eqnarray}
t(s)&=&f_1+f_2\,\sin (b\,s),\nonumber\\
x(s)&=&f_3+f_4\,\cos (2\,b\,s),\nonumber\\
y(s)&=&f_5+f_6\,s,\nonumber\\
z(s)&=&f_7+f_8\,\cos (b\,s),
\label{18}
\end{eqnarray}
where $c_{i}, f_{i}, i=1,\ldots,8$ are arbitrary constants. Taking norm of the tangent vector defined by $g_{\mu\nu}\,\frac{dx^{\mu}}{ds}\,\frac{dx^{\nu}}{ds}$ for the parametric curves (\ref{17}) using the metric (\ref{1}), we get
\begin{eqnarray}
&&g_{\mu\nu}\,\frac{dx^{\mu}}{ds}\,\frac{dx^{\nu}}{ds}\nonumber\\
&=&c_{6}^2-b^2\,c_{2}^2\,\cos^{2} (b\,s)+b^2\,c_{4}^2\,\cos^{2} (b\,s)+b^2\,c_{8}^2\,\sin^{2} (b\,s)\nonumber\\
&-&2\,\alpha_0\,b\,c_2\,c_6\,\cos (b\,s)\,\{c_3+c_4\,\sin (b\,s)\}-\alpha_{0}^2\,c_{6}^2\,[c_3+c_4\,\sin (b\,s)]^2,
\label{19}
\end{eqnarray}
a timelike tangent vector, where we have taken $c_3=5$, $c_4=c_2=1$, $c_6=1$, $c_8=1$, $\alpha_0=1$, $b=1$. We plot a graph of the norm $g_{\mu\nu}\,\frac{dx^{\mu}}{ds}\,\frac{dx^{\nu}}{ds}$ (vertical axis) w. r. t. $s$ (horizontal axis) shown in fig. $1$ (left one).

Similarly, taking norm of the tangent vector for the parametric curves (\ref{18}), one will get
\begin{eqnarray}
&&g_{\mu\nu}\,\frac{dx^{\mu}}{ds}\,\frac{dx^{\nu}}{ds}\nonumber\\
&=&f_{6}^2-b^2\,f_{2}^2\,\cos^{2} (b\,s)+b^2\,f_{8}^2\,\sin^{2} (b\,s)-\alpha_{0}^2\,f_{6}^2\,[f_3+f_4\,\cos (2\,b\,s)]^2\nonumber\\
&-&2\,\alpha_0\,b\,f_2\,f_6\,\cos (b\,s)\,\{f_3+f_4\,\cos (2\,b\,s)\}+4\,b^2\,f_{4}^2\,\sin^{2} (2\,b\,s),
\label{20}
\end{eqnarray}
a timelike tangent vector field, where we have taken $f_3=3$, $f_4=0.1$, $f_2=1$, $f_6=1$, $f_8=1$, $\alpha_0=1$, $b=1$. Ploting a graph of this norm (vertical axis) w. r. t. $s$ (horizontal axis) is shown in fig. $1$ (right one).

\begin{figure}[!tbp]
\includegraphics[width=3.0in,height=2.5in]{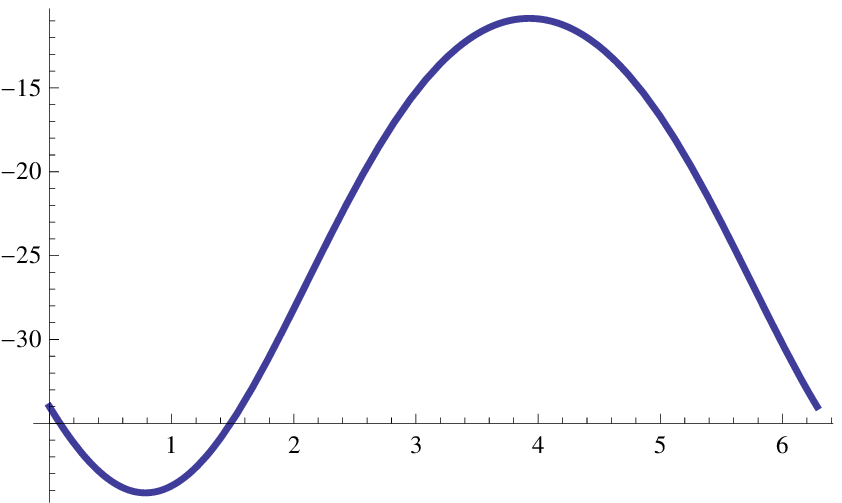}
\hspace{1.0cm}
\includegraphics[width=3.0in,height=2.5in]{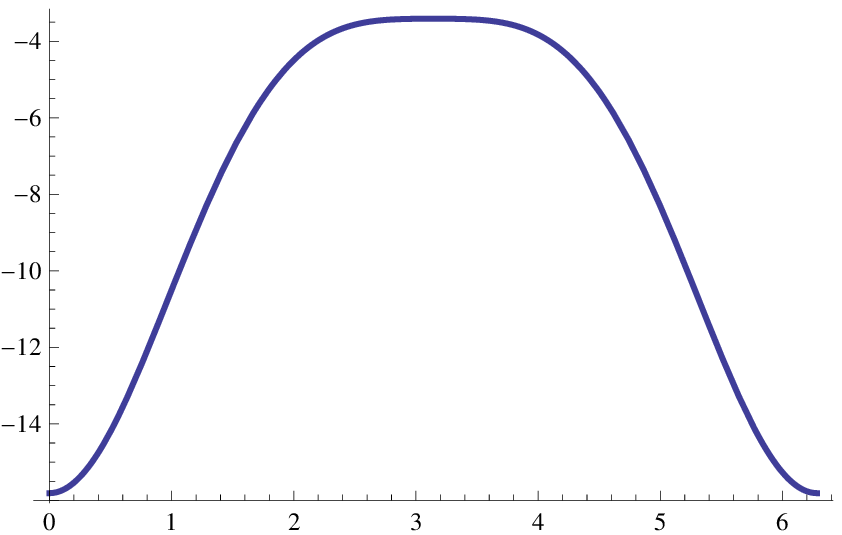}
\caption{Timelike tangent vector}
\end{figure}

Moreover, one can easily show that the above parametric curves are closed in the range $s=0$ to $s=2\,\pi$, {\it i.e.},
\begin{equation}
x^{\mu}(s=0)=x^{\mu}(s=2\,\pi),
\label{closed}
\end{equation}
where $c_6=1=f_6$. Therefore, the parametric curves defined by (\ref{17})--(\ref{18}) being timelike and closed, form closed timelike curves (CTCs).

\section{Null geodesics of the spacetime}

In addition to closed timelike curves, the presented spacetime admits null geodesics which we discuss below.

The spacetime is highly symmetric and admits four Killing vectors in the $(t,x,y)$-subspace. These are $\partial_{t}$, $\partial_{y}$, $y\,\partial_{t}-\frac{1}{\alpha_0}\,\partial_{x}$, $(x^2-y^2)\,\partial_{t}+\frac{2}{\alpha_0}\,(y\,\partial_{x}-x\,\partial_{y})$. To show that third and fourth are the Killing vector, we take the normal form of these given by
\begin{eqnarray}
\chi_{(1)}^{\mu}&=&(y,0,0,0)-\frac{1}{\alpha_0}\,(0,1,0,0)=(y,-\frac{1}{\alpha_0},0,0),\nonumber\\
\chi_{(2)}^{\mu}&=&((x^2-y^2),\frac{2}{\alpha_0}\,y,-\frac{2}{\alpha_0}\,x,0).
\label{kill1}
\end{eqnarray}

The co-variant form of $\chi_{(1)}$ is 
\begin{equation}
\chi_{(1)\,\mu}=(-y, -\frac{1}{\alpha_0}, -\alpha_0\,x\,y, 0 ),
\label{kill2}
\end{equation}
and it satisfies the Killing equation, namely, $\chi_{(1)\,\mu\,;\,\nu}+\chi_{(1)\,\nu\,;\,\mu}=0$. Similarly, one can show that the vector $\chi_{(2)}$ satisfies the Killing equation. Therefore, the vectors, namely, $y\,\partial_{t}-\frac{1}{\alpha_0}\,\partial_{x}$ and $(x^2-y^2)\,\partial_{t}+\frac{2}{\alpha_0}\,(y\,\partial_{x}-x\,\partial_{y})$ are Killing vector. 

The geodesic Lagrangian for the metric (\ref{1}) is
\begin{eqnarray}
L&=&\frac{1}{2}\,g_{\mu\nu}\,\dot{x}^{\mu}\,\dot{x}^{\nu}\nonumber\\
&=&\frac{1}{2}\,[-(\dot{t}+\alpha_0\,x\,\dot{y})^2+\dot{x}^2+\dot{y}^2]
\label{21}
\end{eqnarray}
where we suppress $z$ coordinate and dot represents derivative w. r. t. $\lambda$, an affine parameter. There are two constants of motion corresponding to two cyclic coordinates $t$ and $y$. These are given by
\begin{eqnarray}
\frac{\partial L}{\partial \dot{t}}&=&-E=-(\dot{t}+\alpha_0\,x\,\dot{y})\nonumber\\\Rightarrow 
E&=&(\dot{t}+\alpha_0\,x\,\dot{y}),
\label{hh}
\end{eqnarray}
\begin{eqnarray}
\frac{\partial L}{\partial \dot{y}}&=&K\nonumber\\ \Rightarrow
K&=&\dot{y}-\alpha_0\,x\,(\dot{t}+\alpha_0\,x\,\dot{y})\nonumber\\\Rightarrow
\dot{y}&=&K+\alpha_0\,x\,E.
\label{22}
\end{eqnarray}

One can write eqn. (\ref{hh}) using eqn. (\ref{22}) as
\begin{equation}
\dot{t}=E\,(1-\alpha_0^{2}\,x^2)-\alpha_0\,x\,K.
\label{hh1}
\end{equation}
Thus we have,
\begin{equation}
\Omega(E,K,x)=\frac{dy}{dt}=\frac{\dot{y}}{\dot{t}}=\frac{K+\alpha_0\,x\,E}{E\,(1-\alpha_0^{2}\,x^2)-\alpha_0\,x\,K},
\label{hh2}
\end{equation}
where $\Omega$ is the angular velocity with respect to the stationary observers, {\it i.e.,} observers moving on $t$-lines.

If we let the angular momentum about the z-axis $p_{y}=K=0$, we obtain the angular velocity $\Omega_0$ of a ZAMO (zero angular momentum particle as measured by an observer for whom $t$ is the proper time). This is the angular velocity of the frame dragging \cite{Klein,Mei} and it is given by
\begin{equation}
\Omega_{0}(x)=\frac{\alpha_0\,x}{(1-\alpha_0^{2}\,x^{2})},
\label{hh3}
\end{equation}
which vanishes, {\it i.e.,} $\Omega_0\rightarrow 0$ as $x\rightarrow \pm\,\infty$ and changes sign $\Omega_0>0$ for $x^2<x_0^{2}$ to $\Omega_0<0$ for $x^2>x_0^{2}$, where $x_0^{2}=\frac{1}{\alpha_0^{2}}$.

To show the existence of null geodesic $L=0$ in the spacetime, we first consider the angular momentum is non-zero ($K\neq 0$). From (\ref{21}) using eqn.(\ref{hh})--(\ref{22}) we get
\begin{equation}
\dot{x}^2+\dot{y}^2=(\dot{t}+\alpha_0\,x\,\dot{y})^2\Rightarrow \dot{x}^2=E^2-(K+\alpha_0\,x\,E)^2.
\label{23}
\end{equation}
Writing
\begin{equation}
Y=\frac{K}{E}+\alpha_0\,x,
\label{24}
\end{equation}
we get from eqn. (\ref{23})
\begin{eqnarray}
\dot{Y}&=&\alpha_0\,E\,\sqrt{1-Y^2}\nonumber\\\Rightarrow
\frac{\dot{Y}}{\sqrt{1-Y^2}}&=&\alpha_0\,E.
\label{25}
\end{eqnarray}
The solution of the above equation is 
\begin{equation}
Y(\lambda)=A_1\,\sin (\alpha_0\,E\,\lambda).
\label{26}
\end{equation}
From equation (\ref{24}) we get
\begin{equation}
x(\lambda)=\frac{A_1}{\alpha_0}\,\sin (\alpha_0\,E\,\lambda)-\frac{K}{\alpha_0\,E}.
\label{27}
\end{equation}
Therefore, from eqn. (\ref{22}) we get
\begin{eqnarray}
\dot{y}&=&A_1\,E\,\sin (\alpha_0\,E\,\lambda)\nonumber\\\Rightarrow
y(\lambda)&=&A_2-\frac{A_1}{\alpha_0}\,\cos (\alpha_0\,E\,\lambda),
\label{mm}
\end{eqnarray}
and from eqn. (\ref{hh}) 
\begin{eqnarray}
\dot{t}&=&A_1\,K\,\sin (\alpha_0\,E\,\lambda)-A_1^{2}\,E\,\sin^{2} (\alpha_0\,E\,\lambda)+E\,\nonumber\\\Rightarrow
t(\lambda)&=&A_3+E\,(1-\frac{A_1^{2}}{2})\,\lambda+\frac{A_1^{2}}{4\,\alpha_0}\,\sin (2\,\alpha_0\,E\,\lambda)\nonumber\\
&-&\frac{A_1\,K}{\alpha_0\,E}\,\cos (\alpha_0\,E\,\lambda),
\label{28}
\end{eqnarray}
where $A_i$, $i=1,\ldots,3$ are constants of integration. 

Taking norm of the geodesic eqns. (\ref{27})--(\ref{28}) using the metric (\ref{1}), we get
\begin{equation}
g_{\mu\nu}\,\dot{x}^{\mu}\,\dot{x}^{\nu}=(-1+A_1^{2})\,E^2=0,\quad E\neq 0,
\label{29}
\end{equation}
null geodesics condition provided $A_1=1$. The above null geodesics path are closed in the range of the affine parameter $\lambda=0$ to $\lambda=1$, {\it i.e.,}
\begin{equation}
t(0)=t(1),\quad x(0)=x(1),\quad y(0)=y(1),
\label{ee}
\end{equation}
provided $A_1=\sqrt{2}>1$, where we have chosen $E=\frac{2\,m\,\pi}{\alpha_0}$, $m\in \Re$. As we have seen in (\ref{29}) that null geodesics exist in the spacetime only when $A_1=1$. Therefore, there is no closed null geodesics (CNGs) exist in the spacetime.

For non-zero angular momentum, we have the following null geodesics path 
\begin{eqnarray}
t(\lambda)&=&A_3+\frac{1}{2}\,E\,\lambda+\frac{1}{4\,\alpha_0}\,\sin (2\,\alpha_0\,E\,\lambda)-\frac{K}{\alpha_0\,E}\,\cos (\alpha_0\,E\,\lambda),\nonumber\\
x(\lambda)&=&\frac{1}{\alpha_0}\,\sin (\alpha_0\,E\,\lambda)-\frac{K}{\alpha_0\,E},\nonumber\\
y(\lambda)&=&A_2-\frac{1}{\alpha_0}\,\cos (\alpha_0\,E\,\lambda).
\label{kk1}
\end{eqnarray}
And for zero angular momentum
\begin{eqnarray}
t(\lambda)&=&A_3+\frac{1}{2}\,E\,\lambda+\frac{1}{4\,\alpha_0}\,\sin (2\,\alpha_0\,E\,\lambda),\nonumber\\
x(\lambda)&=&\frac{1}{\alpha_0}\,\sin (\alpha_0\,E\,\lambda),\nonumber\\
y(\lambda)&=&A_2-\frac{1}{\alpha_0}\,\cos (\alpha_0\,E\,\lambda).
\label{kk2}
\end{eqnarray}

Thus the presented spacetime admits null geodesic with the geodesic equations given by (\ref{kk1}) for non-ZAMO and eqn. (\ref{kk2}) for ZAMO (see fig.2, we have chosen $A_3=0$, $\alpha_0=1$, $A_2=1$, $\lambda$ along horizontal axis). 
\begin{figure}[ht]
\centering
\includegraphics[width=3.0in,height=2.5in]{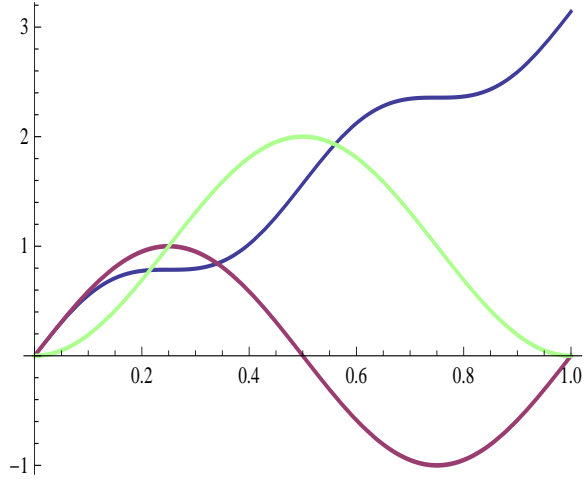}
\caption{Null geodesics for ZAMO : blue--t, violet--x, green--y}
\end{figure}

\section{The Petrov classification of the spacetime}

For classification of the spacetime (\ref{1}), we construct the following set of null tetrad vectors $({\bf k,l,m,\bar{m}})$ \cite{Steph}. They are 
\begin{eqnarray}
k_{\mu}&=&\frac{1}{\sqrt{2}}\,\left (1,0,(1+\alpha_0\,x),0\right),\quad l_{\mu}=\frac{1}{\sqrt{2}}\,\left (1,0,(-1+\alpha_0\,x),0\right),\nonumber\\
m_{\mu}&=&\frac{1}{\sqrt{2}}\,\left(0,1,0,i\right),\quad \bar{m}_{\mu}=\frac{1}{\sqrt{2}}\,\left(0,1,0,-i\right),
\label{30}
\end{eqnarray}
where $i=\sqrt{-1}$. The set of null tetrad above are such that the metric tensor for (\ref{1}) can be expressed as
\begin{equation}
g_{\mu \nu}=-k_{\mu}\,l_{\nu}-l_{\mu}\,k_{\nu}+m_{\mu}\,\bar{m}_{\nu}+\bar{m}_{\mu}\,m_{\nu}.
\label{m1}
\end{equation}
The tetrad vectors (\ref{30}) are null vectors and are orthogonal except for $k_{\mu}\,l^{\mu}=-1$ and $m_{\mu}\,{\bar m}^{\mu}=1$. Using the null tetrads above we calculate the five Weyl scalars of which,
\begin{equation}
\Psi_0=\frac{1}{4}\,\alpha_{0}^2,\quad \Psi_2=\frac{1}{12}\,\alpha_{0}^2,\quad \Psi_4=\frac{1}{4}\,\alpha_{0}^2,
\label{m2}
\end{equation}
are non-vanishing, while others are $\Psi_1=\Psi_3=0$.

One can calculate the Newmann-Penrose spin co-efficients \cite{Steph} using the set of null tetrad vectors (\ref{28}). We find the nonzero spin-coefficients are 
\begin{equation}
\frac{\kappa}{2}=\beta=\alpha=\frac{1}{4\,\sqrt{2}}\,\alpha_0,
\label{31}
\end{equation}
while rest are all equal to zero, where the symbols are same in \cite{Steph}.

An orthonormal tetrad frame ${\bf e}_{(a)}=\{{\bf e}_{(0)},{\bf e}_{(1)},{\bf e}_{(2)},{\bf e}_{(3)}\}$ in terms of  tetrad vectors are
\begin{eqnarray}
{\bf k}&=&\frac{1}{\sqrt{2}}({\bf e}_{(0)}+{\bf e}_{(2)}),\quad {\bf l}=\frac{1}{\sqrt{2}}({\bf e}_{(0)}-{\bf e}_{(2)}),\nonumber\\
{\bf m}&=&\frac{1}{\sqrt{2}}({\bf e}_{(1)}+i\,{\bf e}_{(3)}),\quad {\bf {\bar m}}=\frac{1}{\sqrt{2}}({\bf e}_{(1)}-i\,{\bf e}_{(3)}),
\label{32}
\end{eqnarray}
where ${\bf e}_{(0)}\cdot{\bf e}_{(0)}=-1$ and ${\bf e}_{(i)}\cdot{\bf e}_{(j)}=\delta_{ij}$.

\subsection{The relative motion of the free test particles}

We analyze the effects of the local gravitational fields and the stress-energy tensor terms of the above solutions. For that, we considered the equation of geodesics deviation frame adopted in \cite{Faiz4,Faiz8,Faiz9}. The geodesic equation in terms of orthonormal tetrad are given by
\begin{equation}
\ddot{Z}^{(i)}=-R^{(i)}_{\,(0)(j)(0)}\,Z^{(j)},\quad i,j=1,2,3.
\label{34}
\end{equation}
We set here $Z^{(0)}=0$ such that all test particles are synchronized by the proper time.

From the standard definition of the Weyl tensor we have 
\begin{equation}
R_{(i)(0)(j)(0)}=C_{(i)(0)(j)(0)}-\frac{1}{2}\,[T_{(i)(j)}-\delta_{ij}\,(T_{(0)(0)}+\frac{2}{3}\,T)],
\label{35}
\end{equation}
where $T_{(a)(b)}$ as the stress-energy tensor components and $T=T^{(a)}_{(a)}$.

For the metric (\ref{1}), the only non-vanishing Weyl scalars are given by (\ref{30}) so that
\begin{eqnarray}
C_{(1)(0)(1)(0)}&=&\frac{1}{6}\,\alpha_0^{2},\quad C_{(2)(0)(2)(0)}=\frac{1}{6}\,\alpha_0^{2},\quad C_{(3)(0)(3)(0)}=-\frac{1}{3}\,\alpha_0^{2},\nonumber\\
C_{(1)(2)(1)(2)}&=&\frac{1}{3}\,\alpha_0^{2},\quad C_{(2)(3)(2)(3)}=-\frac{1}{6}\,\alpha_0^{2},\quad C_{(1)(3)(1)(3)}=-\frac{1}{6}\,\alpha_0^{2}.
\label{36}
\end{eqnarray}

One can find out the equations of geodesic deviation (\ref{34}) using (\ref{36}) and the stress-energy tensor (\ref{5}). We find that 
\begin{eqnarray}
\label{37}
\ddot{Z}^{(1)}&=&-R^{(1)}_{\,(0)(j)(0)}\,Z^{(j)}=-\frac{\alpha_0^{2}}{4}\,Z^{(1)},\\
\label{38}
\ddot{Z}^{(2)}&=&-R^{(2)}_{\,(0)(j)(0)}\,Z^{(j)}=-\frac{\alpha_0^{2}}{4}\,Z^{(2)},\\
\label{39}
\ddot{Z}^{(3)}&=&-R^{(3)}_{\,(0)(j)(0)}\,Z^{(j)}=0,
\end{eqnarray}
with the solutions 
\begin{eqnarray}
Z^{(1)}&=&A_{1}\,\cos (\frac{\alpha_0\,\tau}{2})+B_{1}\,\sin (\frac{\alpha_0\,\tau}{2}),\\
Z^{(2)}&=&A_{2}\,\cos (\frac{\alpha_0\,\tau}{2})+B_{2}\,\sin (\frac{\alpha_0\,\tau}{2}),\\
Z^{(3)}&=&A_{3}\,\tau+B_{3},
\end{eqnarray}
where
\begin{equation}
T_{(0)(0)}=\frac{3}{4}\,\alpha_0^{2},\quad T_{(1)(1)}=T_{(2)(2)}=-T_{(3)(3)}=\frac{1}{4}\,\alpha_0^{2},\quad T=-\frac{1}{2}\,\alpha_0^{2},
\label{40}
\end{equation}
and $A_{i}, B_{i}, i=1,2,3$ are arbitrary constants.

\section{Conclusions} 

In this paper, a topologically trivial non-vacuum solution of the Einstein's field equations, was presented. The spacetime is regular everywhere, and free from curvature divergence since the scalar curvature invariants are constant. The metric admits a twisting, shearfree, nonexpanding timelike geodesic congruence. The physical parameters, the energy density $\rho$, the radial pressure $p_{r}$, and the tangential pressure $p_{t}$ are constant satisfy the different energy conditions. The stress-energy tensor anisotropic fluid considered here is a generalization of Equation-of-State (EoS) parameter of perfect fluid $p=\omega\,\rho$ (radiation type), where $\omega=\frac{1}{3}$ is the deviation-free EoS parameter and $\delta=-\frac{2}{3}$ as the deviation parameter from isotropy along the $z$-direction. Additionally, the spacetime admit circular closed timelike curves which appear beyond the null curve, and these timelike curves were found to be linearly stable under small linear perturbation. We shown that the spacetime exhibit the null geodesics curve both for ZAMO and non-ZAMO which are non-closed. Furthermore, we had chosen two set of paramteric curves for the spacetime, and shown that these curves are being closed and timelike, form closed timelike curves. Finally, the physical interpretation of this solution, based on the study of the equation of the geodesics deviation, was presented. It was demonstrated that, this solution depend on the local gravitational fields and the stress-energy terms both of their amplitudes depend on the real number $\alpha_0$.

\end{document}